\documentclass[prb, aps, twocolumn, preprintnumbers, amsmath, amssymb, superscriptaddress]{revtex4-2}

\usepackage[makeroom]{cancel}

\usepackage{amsmath}    
\usepackage{amssymb}
\usepackage{graphicx}   
\usepackage{verbatim}   
\usepackage{color}      
\usepackage{hyperref}   
\usepackage[normalem]{ulem}
\usepackage{natbib}
\usepackage{fixmath}
\usepackage{enumitem}
\usepackage{empheq}
\usepackage[version=3]{mhchem}

\hypersetup{colorlinks,linkcolor=blue,urlcolor=blue,citecolor=blue}

\definecolor{darkgreen}{rgb}{0,0.5,0}
\definecolor{purple}{rgb}{0.35,0,0.35}
\definecolor{orange}{rgb}{1,0.5,0}
\definecolor{darkred}{rgb}{.7,0,0}
\definecolor{darkblue}{rgb}{0,0,.3}
\definecolor{grey}{rgb}{.6,.6,.6}
\definecolor{dimgreen}{rgb}{0.2,0.6,0.1}

\newcommand{\average}[1]{\langle #1 \rangle}

\newcommand{\cD}{{\cal D}}

\newcommand{\cL}{\cal L}
\newcommand{\cT}{\cal T}
\newcommand{\cP}{\cal P}
\newcommand{\cPT}{\cal PT}

\definecolor{darkgreen}{rgb}{0,0.5,0}

\begin{document}

\title{\texorpdfstring{$\mathcal{PT}$}{PT}-symmetry phase transition in a Bose-Hubbard model with localized gain and loss}

\author{C\u{a}t\u{a}lin Pa\c{s}cu Moca}
\affiliation{Department of Theoretical Physics, Institute of Physics, Budapest University of Technology and Economics, Műegyetem rkp.~3, H-1111 Budapest, Hungary}
\affiliation{Department  of  Physics,  University  of  Oradea,  410087,  Oradea,  Romania}
\author{Doru Sticlet}
\email{doru.sticlet@itim-cj.ro}
\affiliation{National Institute for R\&D of Isotopic and Molecular Technologies, 67-103 Donat, 400293 Cluj-Napoca, Romania}
\author{Bal\'azs D\'ora}
\affiliation{Department of Theoretical Physics, Institute of Physics, Budapest University of Technology and Economics, Műegyetem rkp.~3, H-1111 Budapest, Hungary}
\affiliation{MTA-BME Lend\"ulet Topology and Correlation Research Group, Budapest University of Technology and Economics, Műegyetem rkp.~3, H-1111 
Budapest, Hungary}
\author{Gergely Zar\' and}
\affiliation{Department of Theoretical Physics, Institute of Physics, Budapest University of Technology and Economics, Műegyetem rkp.~3, H-1111 Budapest, Hungary}
\affiliation{MTA-BME Quantum Dynamics and Correlations Research Group, Budapest University of Technology and Economics, Műegyetem rkp.~3, H-1111 Budapest, 
Hungary}


\begin{abstract}
We study the dissipative dynamics of a one-dimensional bosonic system described in terms of the bipartite Bose-Hubbard model with alternating gain and loss. This model exhibits the $\cP\cT$ symmetry under some specific conditions and features a $\cP\cT$-symmetry phase transition. It is characterized by an order parameter corresponding to the population imbalance between even and odd sites, similar to the continuous phase transitions in the Hermitian realm. 
In the noninteracting limit, we solve the problem exactly and compute the parameter dependence of the order parameter. The interacting limit is addressed at the mean-field level, which allows us to construct the phase diagram for the model. 
We find that  both the interaction and dissipation rates induce $\cPT$-symmetry breaking.
On the other hand, periodic modulation of the dissipative coupling in time stabilizes the $\cPT$-symmetric regime.
Our findings are corroborated numerically on a tight-binding chain with gain and loss.
\end{abstract}
\maketitle

\section{Introduction}

In classical mechanics and classical field theory, symmetry transformations play a prominent role because they result in conservation laws, which typically facilitate the solution of physical problems~\cite{Noether1918}. 
They have a similar importance in quantum physics, where they, for instance, enable the development of selection rules for quantum transitions, and occasionally allow even the precise determination of spectra using solely group theoretic techniques~\cite{Tinkham2003}. 
Symmetries become even more crucial in quantum field theories where they typically serve as the very foundation of their formulation by accounting for the observed or postulated conservation law~\cite{Altland2010}. 
Recently, within the context of non-Hermitian quantum mechanics and open quantum systems,
two particular symmetries became relevant: the parity or the spatial inversion $\cP$ and the time-reversal symmetry $\cT$. 
If the action of the combined $\cPT$ operators on the Hamiltonian satisfies the relation ${\cP\cT} H({\cP\cT})^{-1}=H $, then the Hamiltonian is said to 
be $\cP\cT$ symmetric~\cite{bender98,Bender2007,ptreview}. 

Systems with $\cP\cT$ symmetry may be non-Hermitian and feature real spectra when the eigenstates are also $\cPT$ invariant ~\cite{Mostafazadeh2002,Bender2007}.
In general, such systems display exceptional points (EP) in the parameter space where the $\cP\cT$ symmetry breaks, the eigenspectrum becomes degenerate, and the 
eigenstates coalesce~\cite{Kato1995,Heiss.2004, heiss}. 
At such points, the system features a real-to-complex transition in the energy spectrum. 
Besides the  EPs, non-Hermitian quantum mechanics provides several other unique phenomena, absent in the Hermitian realm, such as unidirectional invisibility~\cite{Lin2011}, non-Bloch oscillations~\cite{Longhi.2019}, and skin effect~\cite{Yao.2018,Kunst2018,Song2019},
to mention a few. 

Non-Hermiticity is perfectly suited to analyze systems with dissipation~\cite{Ashida2020}. 
In this context, a non-Hermitian ``effective Hamiltonian'' can represent the quantum dissipation process at the mean-field level when a quantum system coupled to a surrounding environment or bath is under continuous monitoring~\cite{Jacobs2006,Wiseman2009}, and therefore 
the recycled dissipative term in the full Lindbladian evolution of the reduced density matrix becomes irrelevant~\cite{Daley2014}.

It is therefore natural to  extend these symmetry considerations to Liouville operators and investigate the dissipative dynamics of an open system that features the $\cP\cT$ symmetry~\cite{Huber2020,Huber2020a,Nakanishi2022}. 

In general, the state of an open system is described by the density matrix $\rho$ whose time evolution is given by the Lindblad equation~\cite{Lindblad1976,Gorini1976}
\begin{equation}
i \frac{d\rho(t)}{dt} = {\cL}[\rho(t)] = [H, \rho(t)] +{\cD} [\rho(t)] \,,
\end{equation}
where $\cL$ is the Lindbladian, $H$ represents the system Hamiltonian, and $\cD$ denotes the dissipative part of $\cL$, 
\begin{equation}
{\cD}[\rho(t)] = i\sum_{\mu} \Gamma_\mu \big( 2F_\mu \rho F_\mu^\dagger -\{F_\mu^\dagger F_\mu, \rho \}\big)\,,
\end{equation}
It contains a set of dissipative operators $F_{\mu}$ that model interaction between system and environment. 
The coupling of the system to the external environment is described by the rates $\Gamma_\mu$. 
Here we do not impose the whole Lindbladian to be $\cPT$ symmetric, 
but the $\cPT$ symmetry applies to the Hamiltonian and to the dissipative part of the Lindbladian separately, a condition that implies that the $\cPT$ symmetry is satisfied by the expectation values as 
well~\cite{Huber2020,Nakanishi2022}. 
Therefore, we consider a generic open system to be $\cP\cT$ symmetric if the Lindblad operator has the property
\begin{equation}
{\cL }[{\cPT} H ({\cPT})^{-1}, \{ {\cPT}F_{\mu}({\cPT})^{-1}\} ] = {\cL}[H, \{F_\mu\}].
\label{eq:ptsym}
\end{equation}
It can be demonstrated that only open bosonic models are $\cP\cT$ symmetric, whereas in fermionic systems, the anticommutation relations prevent this from being the case~\cite{Huber2020}.

In the present work, we focus on the one-dimensional bipartite Bose-Hubbard model with alternating gain and loss, which, under some particular circumstances, exhibits the $\cPT$ symmetry and displays a $\cPT$-symmetry phase transition, which is characterized by an order parameter similar to that of continuous phase transitions in regular Hermitian systems~\cite{Goldenfeld1972}.
We solve exactly the dynamics in the noninteracting limit and extract the scaling exponents for the order parameter. 
The interaction limit has also been handled at the mean-field level, which allows us to construct the phase diagram for the model.
We find that both dissipation and interaction break $\cPT$ symmetry.
We must always keep in mind that the Markovian evolution~\cite{Breuer2007} is generally associated with a large, $T\to \infty$, temperature, 
and therefore the corresponding $\cPT$-symmetry phase transition occurs at large temperatures, in a nonequilibrium steady state, in contrast 
to the regular continuous phase transitions that typically take place at low temperatures~\cite{Sondhi.1997} and change the ground state of the system.

In the absence of dissipation, the model has been extensively studied~\cite{Fisher1989,Freericks1994, Kuhner1998,Cazalilla.2011}.
The ground state of the system realizes an interaction-driven Mott insulating phase for strong interactions, while
for weaker interactions it remains a superfluid. 
This was originally demonstrated in a landmark study~\cite{Fisher1989}, which examined the universality class of the model 
and offered a mean-field 
solution.
The one-dimensional (1D)  Bose-Hubbard Hamiltonian, [introduced below in \eqref{eq:Ham_1}], cannot be solved by the Bethe ansatz, and knowledge of its characteristics is only partial in 1D, unlike the fermionic counterpart. This requires the use of other available tools such 
as quantum Monte Carlo~\cite{Batrouni.1990} to address its dynamics, or to observe the 
destruction of the superfluidity phase by disorder~\cite{Scalettar.1991} and 
the formation of the Bose glass phase. Entanglement spreading and evolution of correlations have also been investigated 
recently~\cite{Lauchli.2008}.

The Bose-Hubbard model has been implemented for the first time in cold atomic settings  in Ref.~\cite{Jakschh1998}, enabling 
quantum simulation and an examination of the correlations and dynamics of the model. 
Many other Hamiltonian realizations have since been attained (see, e.g., Refs.~\cite{Paredes.2004, Morsch.2006}, or \cite{Lewenstein.2007}).
The dissipative form of the Bose-Hubbard model, which appears to prefer the Mott insulating state, may now be engineered with the advent of 
improved experimental methods~\cite{Takafumi.2017}. 
Furthermore, utilizing \ce{^{174}Yb} atoms in an optical superlattice with one-body atom loss as dissipation in Ref.~\cite{Takasu.2020}, the 
realization of a $\cPT$-symmetric model has been reported,  demonstrating that our idea is well within experimental reach.

The rest of the paper is organized as follows. 
In Sec.~\ref{sec:Model}, we introduce the model and derive the general equations for the two-point correlators. 
In Sec.~\ref{sec:U=0}, we solve the noninteracting limit and compute the order parameter. 
In Sec.~\ref{sec:U_ne_0}, we present the mean-field solution, construct the phase diagram, and analyze the case of periodic driving, while in Sec.~\ref{sec:Conclusions} we summarize the main findings.

\section{Bose-Hubbard model with alternating gain and loss}\label{sec:Model}
\begin{figure}[t!]
	\includegraphics[width=0.9\columnwidth]{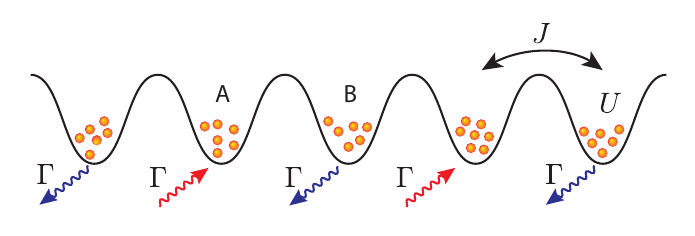}
	\caption{Schematic representation of the bipartite Bose-Hubbard model coupled in an alternating way to the environment. 
	At sites $A$, the particles are injected into the systems while at the sites $B$, the system is losing particles. 
	The hopping integral between two nearest neighbor sites is denoted by $J$, while $U$ represents the on-site Coulomb interaction, 
	and $\Gamma$ is the coupling strength between system and environment and is considered to be constant for all sites.}
	\label{fig:sketch_model}
\end{figure}
We assume an interacting gas of bosonic particles on a one dimensional lattice with alternating particle loss 
and gain as displayed in Fig.~\ref{fig:sketch_model}. We model it by the  Bose-Hubbard Hamiltonian~\cite{Fisher1989} 
\begin{equation}
H = J\sum_{n} \big (b_n^\dagger b_{n+1} +h.c. \big) +\frac{U}{2} \sum_{n} b_n^\dagger b_n^\dagger b_n b_n, 
\label{eq:Ham_1}
\end{equation}
where $J$ represents the nearest neighbor hopping, $U$ is the on-site repulsion energy, and $b_n^{(\dagger)}$ are the typical annihilation (creation) bosonic operators at site $n$. 
In what follows, we use natural units $\hbar=1$, and $J$ represents the unit of energy, such that time is measured in units of $J^{-1}$ and dissipation rates in units of $J$.

We immediately notice that the Hamiltonian~\eqref{eq:Ham_1} belongs to the $\cPT$-symmetric models.
The dissipative environment must be built in a way that Eq.~\eqref{eq:ptsym} is satisfied.
For that, we assume an alternating, single-particle gain-and-loss environment 
described by the jump operators $b^\dagger_n$ and $b_n$, but with a constant 
system-environment coupling $\Gamma$ in both the gain and loss channels. 
Applying the $\cPT$-symmetry operation converts the gain portion of the Lindbladian into a loss and vice versa, and therefore the $\cPT$ symmetry is thus satisfied for the dissipative part of the Lindbladian as well.

Because of the alternating sites with gain and loss, the lattice can be viewed as being bipartite~\cite{Asb0th.2016}, with two sites ($A$ and $B$) per unit cell. 
At sites labeled $A$ (even sites), the system loses particles while at sites $B$ (odd sites), bosons are injected into the system. 
Using this convention, we rewrite the Hamiltonian~\eqref{eq:Ham_1} by explicitly using the sublattice labels as well, i.e.,  $b_{n}\to b_{j, \alpha}$, where $j$ labels the unit cell, while $\alpha=\{A,B\}$, the site inside the unit cell. 
In this new basis, the Hamiltonian is  
\begin{gather}
H = J\sum_{j=-\infty}^{\infty}\big\{ b^\dagger_{j,A}\big( b_{j,B}+b_{j-1, B}  \big) +{\rm H.c.}\big\}\nonumber \\
{}+\frac{U}{2} \sum_j \sum_{\alpha =A,B}
b^\dagger_{j,\alpha} b^\dagger_{j,\alpha} b_{j, \alpha} b_{j,\alpha}\, .
\end{gather}

To study the $\cPT$-symmetry breaking transition, we introduce the order parameter as the population imbalance between the even and odd sites,
%
\begin{equation}
\Delta = \frac{ |n_A- n_B| }{n_A+n_B},
\end{equation}
where $n_{\alpha}=\average{c^\dagger_{j, \alpha} c_{j, \alpha}}$ is the average occupation on the sublattice $\alpha$. 
For an infinite lattice (or when employing periodic boundary conditions), the model remains translation invariant, suggesting that the 
average occupation on each sublattice is position independent.

We begin our analysis by looking into the time-dependent, two-point correlation function~\cite{Krapivsky2020}, whose diagonal components represents the average occupation. 
The two-point correlator is defined as 
\begin{equation}
\sigma_{ij}^{\alpha\beta} (t)= \average{b^\dagger_{i\alpha} b_{j\beta}}=\textrm{Tr}\{ \rho(t)b^\dagger_{i\alpha} b_{j\beta} \}\,.
\label{eq:sigma}
\end{equation}
The time dependence of correlators is determined using the equation of motion method~\cite{Mahan2000}.
In general, this approach generates higher order correlators along the way, but in the case of a noninteracting problem, corresponding to $U=0$, it generates a closed set of equations for $\sigma_{ij}^{\alpha\beta}(t)$ that are solved exactly. 
On the other hand, the interacting model, corresponding 
to $U\ne 0$ cannot be solved exactly because of these nonzero higher and higher moments.
However, at the mean-field level, associated with a large number of particles in the system, $N\to \infty$, 
a good approximation consists in neglecting the variances of four-point correlation functions and higher, 
and eventually close the system of equations.
The time dependence is evaluated by taking the derivative of Eq.~\eqref{eq:sigma} and using the Heisenberg equation of motion for the operators. 
After performing this truncation, at the mean-field level, the two-point correlator satisfies the differential equation

\begin{eqnarray}
\dot\sigma_{i\,j}^{\alpha\beta} (t)&=& iJ\big ( \sigma_{i+1\, j}^{\bar\alpha\beta}+ \sigma_{i-1\, j}^{\bar\alpha\beta} -
\sigma_{i\, j+1}^{\alpha\bar\beta} -\sigma_{i\, j-1}^{\alpha\bar\beta} \big)\nonumber\\
&&{}+i U\big( \sigma_{i\, i}^{\alpha\alpha} \sigma_{i\, j}^{\alpha\beta} -\sigma_{i\, j}^{\alpha\beta} \sigma_{j\, j}^{\beta\beta}  \big)\nonumber\\
&&{}+\Gamma \sigma_{i\, j}^{\alpha\beta} \big( \delta^{\alpha A}+\delta^{\beta A}- \delta^{\alpha B} -\delta^{\beta B} \big)\label{eq:sigma_der}\\
&&{}+2\Gamma\delta^{\alpha A}\delta^{\beta A} \delta_{ij}\nonumber,
\end{eqnarray}
where we used the convention that $\bar A=B$. In deriving Eq.~\eqref{eq:sigma_der}, we have performed the approximation
$\average{b^\dagger_{i\alpha}b_{i\alpha} b^\dagger_{i\alpha} b_{j \beta}}\approx \sigma_{i\, i}^{\alpha\alpha} \sigma_{i\, j}^{\alpha\beta}$ and neglected the covariance $\Delta = \average{b^\dagger_{i\alpha}b_{i\alpha} b^\dagger_{i\alpha} b_{j \beta}} -  \sigma_{i\, i}^{\alpha\alpha} \sigma_{i\, j}^{\alpha\beta}$. 

Given that the covariance scales linearly with the number of bosons, $\Delta \sim N$, while the product scales quadratically, $\sigma_{i\, 
i}^{\alpha\alpha} \sigma_{i\, j}^{\alpha\beta}\sim N^2 $, this is a realistic approximation in the limit of a large occupancy number per site. At the level of 
mean-field approximation that we discuss here, the correction made by explicitly incorporating the covariances is of the order of ${\cal O}(1/N)$, so we disregard it.

We first analyze the homogeneous equation of motion~\eqref{eq:sigma_der} without including the last term in Eq.~\eqref{eq:sigma_der}, which 
represents the source term. By mapping it to a discrete $\cPT$-symmetric version of a non-Hermitian Gross-Pitaevskii equation (GPE) and using the factorization ansatz for the homogeneous part of the correlator $\sigma_{i\, j, 0}^{\alpha\beta}(t) = \psi^*_{\alpha, i} (t) \psi_{\beta, j} (t) $ we get
\footnote{The homogeneous correlator $\sigma_{i\, j, 0}^{\alpha\beta}(t)$ satisfies Eq.~\eqref{eq:sigma_der} but without the source term $2\Gamma\delta^{\alpha A}\delta^{\beta B}\delta_{ij}$ included.}
\begin{equation}\label{eq:psi_general}
\begin{aligned}
\dot \psi_{A, j} (t) &= iJ \big( \psi_{B, j-1}+\psi_{B, j}  \big) +i U |\psi_{A, j}|^2 \psi_{A, j}+\Gamma \psi_{A, j},\\
\dot \psi_{B, j} (t) &= iJ \big( \psi_{A, j}+\psi_{A, j+1}  \big) +i U |\psi_{B, j}|^2 \psi_{B, j}-\Gamma \psi_{B, j}.
\end{aligned}
\end{equation}
Then, the full solution for the correlator, including the source term, is obtained as
\begin{equation}\label{corr_mat}
\sigma_{i\, j}^{\alpha\beta}=2\Gamma \int_0^t d\tau\, \psi^*_{\alpha, i}(\tau )\psi_{\beta, j}(\tau).
\end{equation}
From this, the  number of particles along the chain is evaluated as
\begin{equation}
\average{n_{i,\alpha}(t)} = 2\Gamma \int_0^t d\tau |\psi_{\alpha, i}(\tau)|^2.
\label{eq:n_i}
\end{equation}
The wave functions $\psi_{\alpha, j}(t)$ exhibit oscillating behavior in the $\cPT$-symmetric phase, but change to exponential growth with a certain effective rate in the $\cPT$-broken phase, as we will examine next. 
As a result, in the $\cPT$-symmetric phase, apart from some superimposed oscillations, the average occupancy grows linearly with time, while in the $\cPT$-broken regime, it increases exponentially.

\section{Noninteracting limit, \texorpdfstring{$U=0$}{U=0}}\label{sec:U=0}
In the noninteracting limit, the cubic terms in Eq.~\eqref{eq:psi_general} 
drop out and the linear set of equations is solved exactly by performing a Laplace transformation with respect to time, followed by a
Fourier transformation to momentum space~\cite{Krapivsky2020}. 
Introducing the two transformations
\begin{eqnarray}
\hat \psi_{\alpha, j}(s)&=&\int_0^\infty dt e^{-s t} \psi_{\alpha, j}(t),\nonumber \\
 \psi_{\alpha}(s,q)&=&\sum_{j}\hat \psi_{\alpha, j}(s) e^{-i q j}, 
\end{eqnarray}
the differential set of equations~\eqref{eq:psi_general} become a set of linear equations in Fourier space
\begin{eqnarray}
(s-\Gamma)\psi_{A}(s,q) - 2iJ e^{-i q/2}\cos{\frac q2} \psi_{B}(s, q) &=& 2\pi \delta(q),\nonumber\\
- 2iJ e^{i q/2}\cos{\frac q2} \psi_{A}(s, q)+(s+\Gamma)\psi_{B}(s, q) &=& 0,\nonumber
\end{eqnarray}
with the solution for the two components:
\begin{eqnarray}
\psi_A(s,q) &= &{2\pi (s+\Gamma)\delta(q)\over s^2-\Gamma^2+4J^2\cos^2 {q\over 2}}, \\
\psi_B(s,q) &= &{4i \pi Je^{i \frac q2}\cos{\frac q2} \delta(q)\over s^2-\Gamma^2+4J^2\cos^2 {q\over 2}}.
\end{eqnarray}

\begin{figure}[t!]
	\includegraphics[width=0.6\columnwidth]{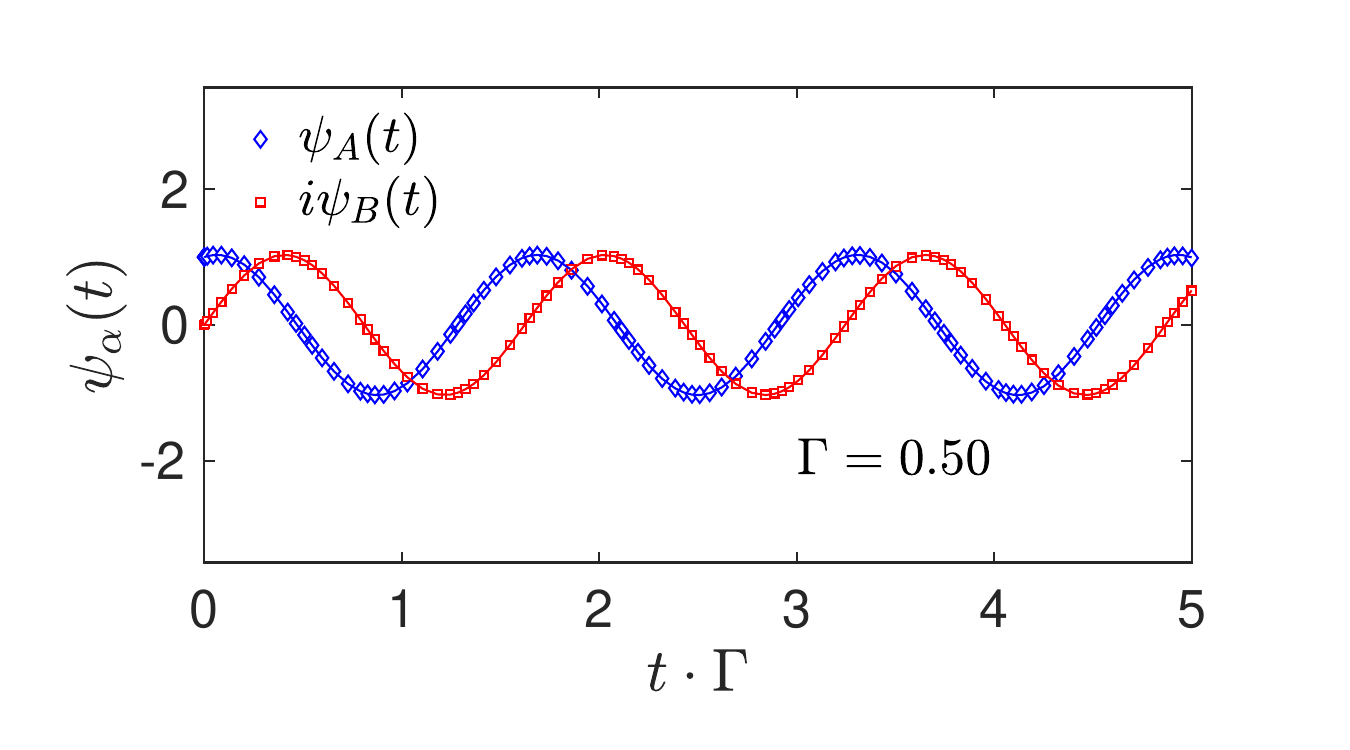}
	\includegraphics[width=0.33\columnwidth]{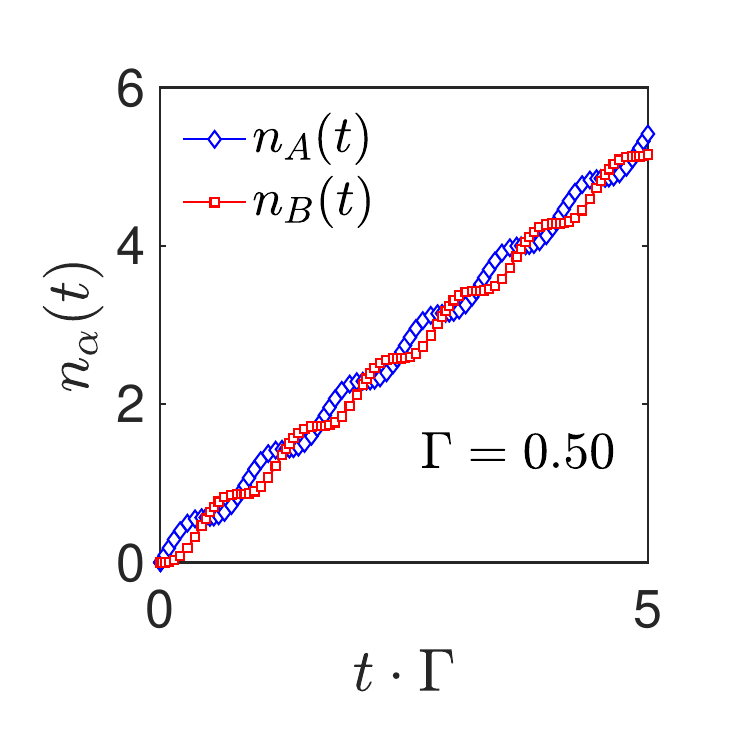}	
	\includegraphics[width=0.6\columnwidth]{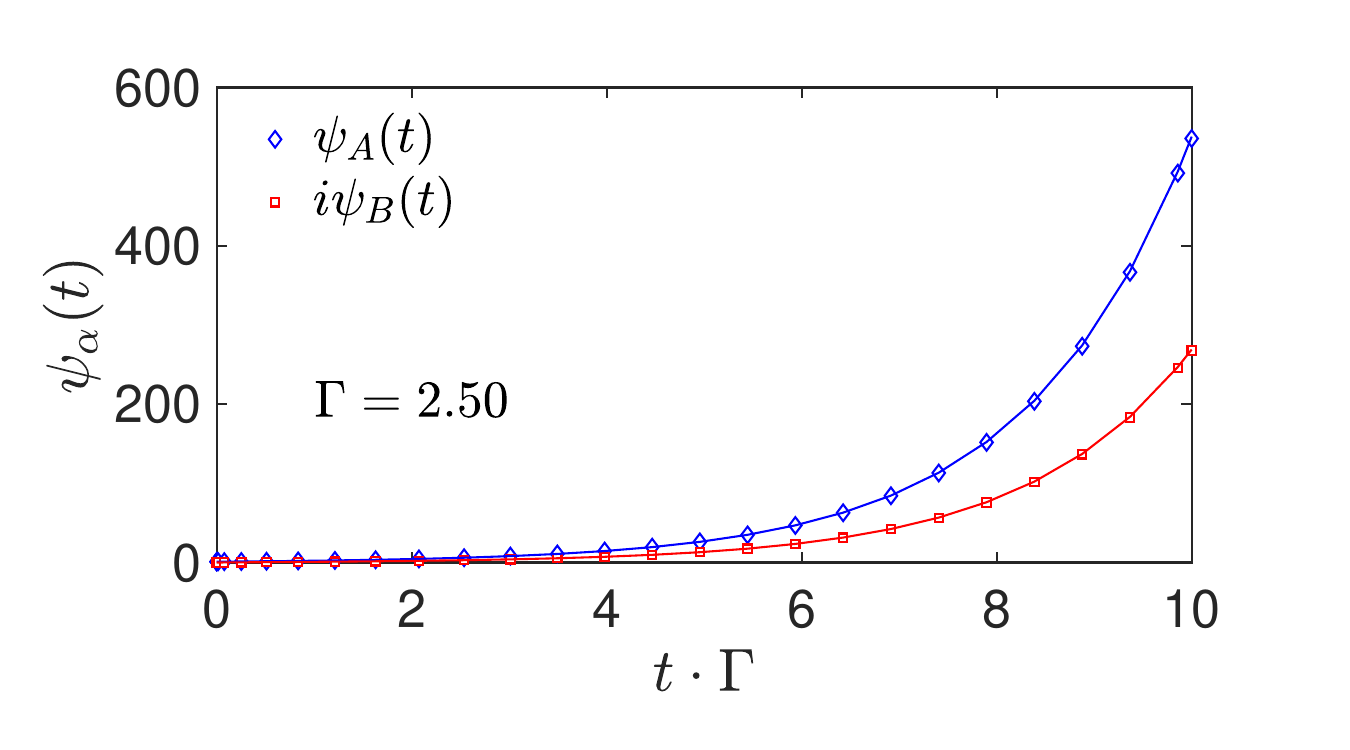}
	\includegraphics[width=0.33\columnwidth]{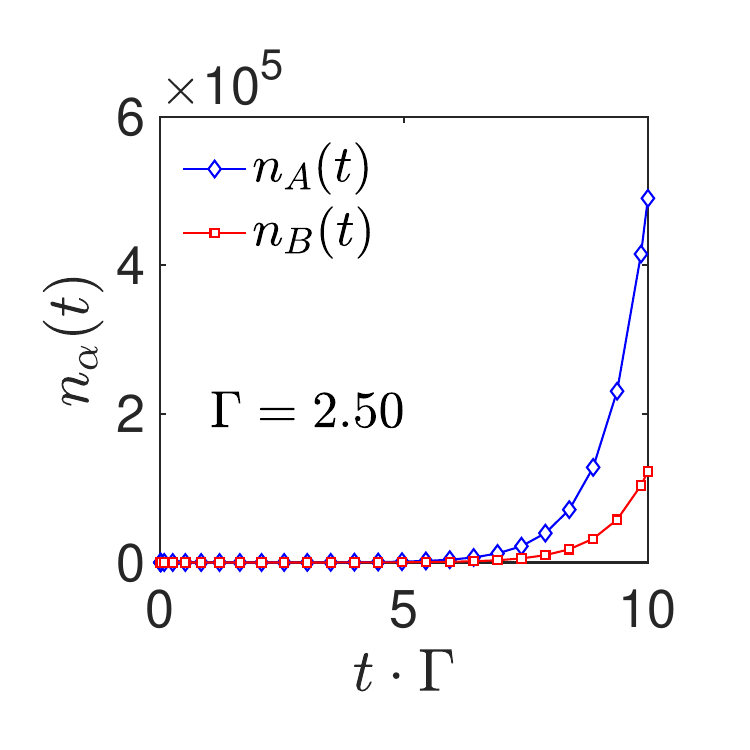}	
	\caption{Comparison between the analytical results for $\psi_{\alpha}(t)$ given in Eqs.~\eqref{eq:S_AB_cos} and \eqref{eq:S_AB_cosh} (solid lines) 
	and the numerical solution obtained by integrating numerically the homogeneous equation Eq.~\eqref{eq:psi_general} (symbols).
	The upper panel displays the periodic time dependence of $\psi_{\alpha}(t)$ in the $\cPT$-symmetric regime with $\Gamma<\Gamma_C$, while the lower panel shows the behavior in the $\cPT$-broken phase where the $\psi_{\alpha}(t)$ diverges in the long-time limit when $\Gamma>\Gamma_C$.	
	The right panels display the occupation of each site $n_{\alpha}(t)$ as a function of time.
	}
	\label{fig:comparison_S_AB}
\end{figure}

\begin{figure}[t!]
	\includegraphics[width=0.9\columnwidth]{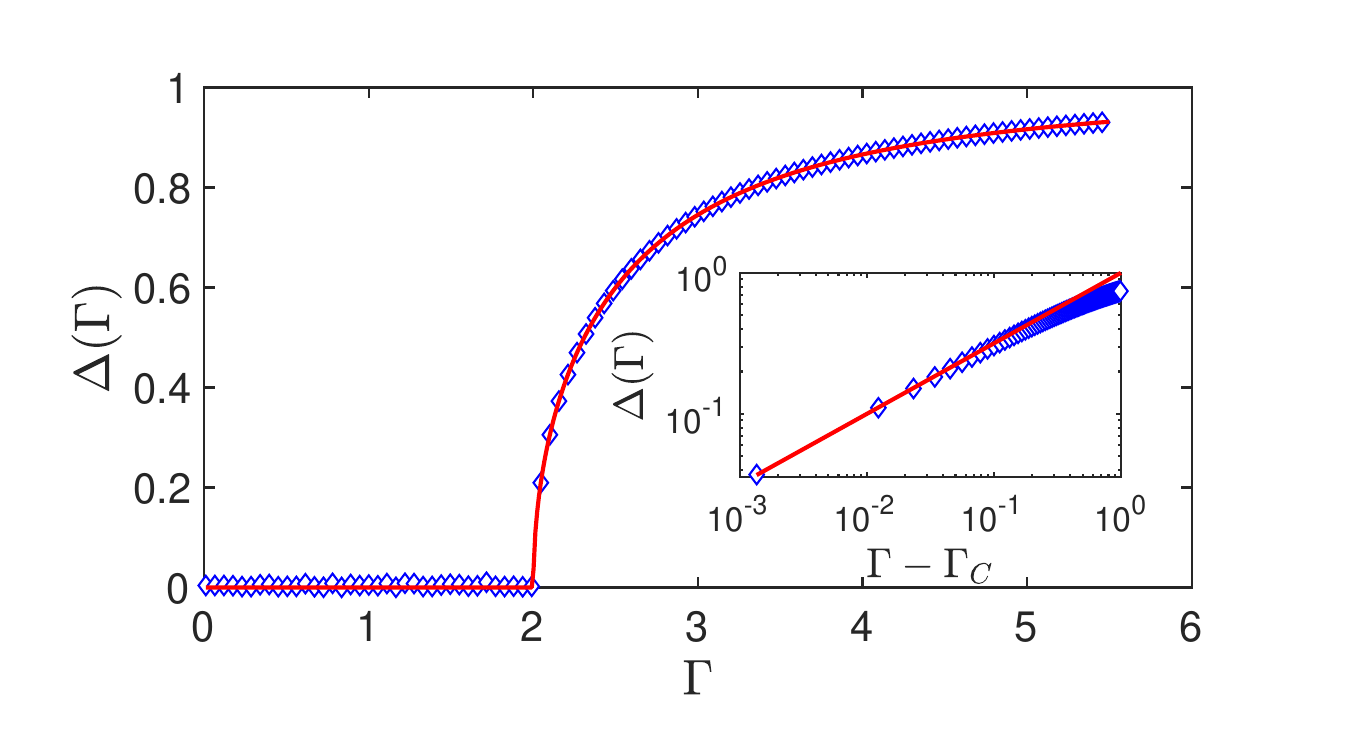}
	\caption{Evolution of the order parameter $\Delta$  as function of $\Gamma$ across the $\cPT$-phase transition. The symbols are the numerically computed values, while
		the solid line represents the analytical result in Eq.~\eqref{eq:Delta}. (Inset) The scaling of $\Delta$ close to the transition point
		capturing the mean field critical behavior. The red solid line corresponds to the power-law behavior $\sim (\Gamma-\Gamma_C)^{0.5}$. }
	\label{fig:comparison_Delta}
\end{figure}

\begin{figure}[t!]
\includegraphics[width=\columnwidth]{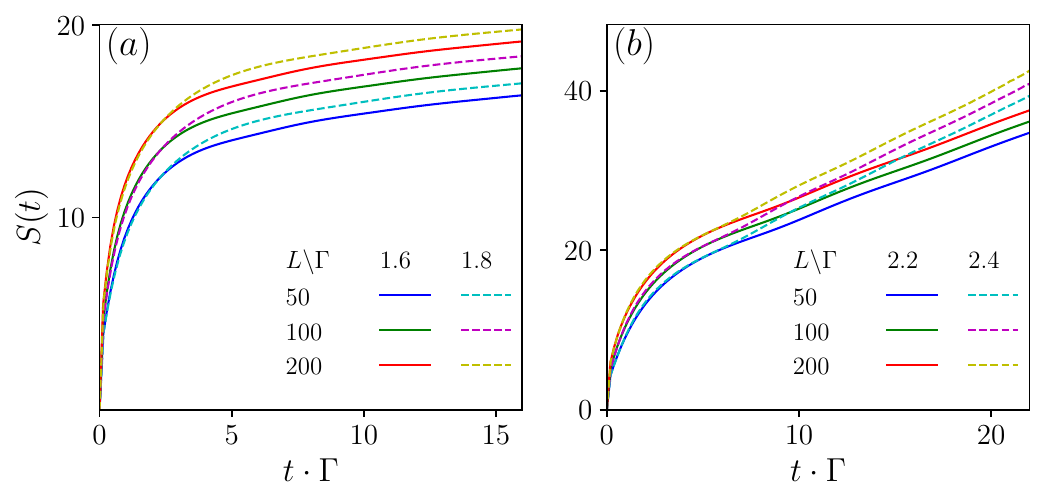}
\caption{Thermodynamic entropy  as a function of time.
(a) In the $\cPT$-symmetric phase, $S(t)$ has logarithmic growth as function of time, $S(t)\propto \log t$. 
(b) In the $\cPT$-broken phase, $S(t)$  grows linearly in time $S(t)\propto t$. 
In both cases, we present results for three system sizes $L$ (the number of unit cells) and two  rates $\Gamma$.}
\label{fig:entropy}
\end{figure}

Performing the inverse Fourier transform, followed by the Laplace transform, we obtain an exact expression for the time-dependent functions $\psi_{\alpha, j}(t)$. 
Their expressions are site independent, and because of that, in what follows, we drop the site label $j$, and keep only the sublattice label $\alpha$. 

The nature of the solution changes with respect to the value of $\Gamma$. For $\Gamma >\Gamma_C$ with critical coupling $\Gamma_C = 2J$, we get the following expressions:
\begin{eqnarray}
\psi_A(t) &=& \cosh (\Gamma^>_{\rm eff}\,t) +{\Gamma\over \Gamma^>_{\rm eff} } \sinh (\Gamma^>_{\rm eff} t),\nonumber \\
\psi_B(t) &=& {i\Gamma_C\over \Gamma^>_{\rm eff}} \sinh (\Gamma^>_{\rm eff} t),
\label{eq:S_AB_cosh}
\end{eqnarray}
where $\Gamma^>_{\rm eff}=\sqrt{\Gamma^2-\Gamma_C^2}$.
In contrast, the functions are evaluated for $\Gamma<\Gamma_C$ as
\begin{eqnarray}
\psi_A(t) &= &\cos (\Gamma^<_{\rm eff}\,t) +{\Gamma\over \Gamma^<_{\rm eff}} \sin (\Gamma^<_{\rm eff} t),\nonumber\\
\psi_B(t) &= &{i\Gamma_C\over \Gamma^<_{\rm eff}} \sin (\Gamma^<_{\rm eff} t),
\label{eq:S_AB_cos}
\end{eqnarray}
with $\Gamma^<_{\rm eff}=\sqrt{\Gamma_C^2-\Gamma^2}$.
Notice that solution~\eqref{eq:S_AB_cos} and~\eqref{eq:S_AB_cosh} are either purely real or imaginary when $U=0$. 
As we will see next, at finite $U$ the solutions are always complex and periodic, irrespective of $U$ or $\Gamma$ strength.  
To check the accuracy of our solution, we solved the set of equations~\eqref{eq:psi_general}  numerically as well. 
A comparison between the two approaches is presented in Fig.~\ref{fig:comparison_S_AB}
for the average occupation as function of time, obtained by integrating  Eq.~\eqref{eq:n_i}.
In the $\cPT$-symmetric regime, the bounded, oscillating behavior of $\psi_{\alpha}(t)$ as obtained in Eq.~\eqref{eq:S_AB_cos} leads to a linear increase in time of the average occupation 
\begin{equation}
\average{n_\alpha(t)}\approx {\Gamma \over \Gamma^<_{\rm eff}}\, t \label{eq:n_t}
\end{equation} 
with a rate $\Gamma/\Gamma^<_{\rm eff}$.
Superimposed, it exhibits an oscillating behavior with a frequency  $\sim \Gamma^<_{\rm eff}$. 
On the other hand, in the $\cPT$-broken regime, integrating Eq.~\eqref{eq:S_AB_cosh} reveals an exponential increase in the occupation as function of time.
Both behaviors are displayed in the right columns in Fig.~\ref{fig:comparison_S_AB}. At the critical point $\Gamma=\Gamma_c$, $\langle n_\alpha(t)\rangle \sim \frac23 \Gamma_C^3 t^3$.

We observe that the lattice model reduces to a double-well potential problem when periodic boundary conditions (PBC) are applied, which provides a link to non-Hermitian physics~\cite{Graefe2012,Kreibich2013,Dast2014,Dizdarevic2018,Zhang2021}. 
Therefore, without loss of generality, Eq.~\eqref{eq:psi_general} for the lattice model is recast as a Schr\" odinger equation for a simpler, 
non-Hermitian two-level system of the form
\begin{equation}
i{d\over dt}
\left (
\begin{array}{c}
\psi_A(t)\\
\psi_B(t)
\end{array}
\right ) =
H^{\rm eff}_0
 \left (
\begin{array}{c}
\psi_A(t)\\
\psi_B(t)
\end{array}
\right ),
\label{eq:Schro}
\end{equation}
with symmetric gain and loss at the two sites. Furthermore, 
we associate the quantities $\psi_\alpha(t)$ in Eq.~\eqref{eq:psi_general} with the wave functions describing the effective non-Hermitian Hamiltonian
\begin{equation}
H^{\rm eff}_0=\left (
\begin{array}{cc}
i \Gamma & -2J\\
-2J & -i\Gamma 
\end{array}
\right ).\label{eq:H0}
\end{equation}
The Hamiltonian $H^{\rm eff}_0$ for the simplified model remains $\cPT$ symmetric 
with respect to discrete parity operator ${\cP}=\sigma_x$, the regular Pauli matrix, 
and the time-reversal operator, the complex conjugate operator ${\cT}:i\to -i$, and 
furthermore presents an EP at $\Gamma =\Gamma_C$. 
\begin{figure}[t!]
	\includegraphics[width=0.6\columnwidth]{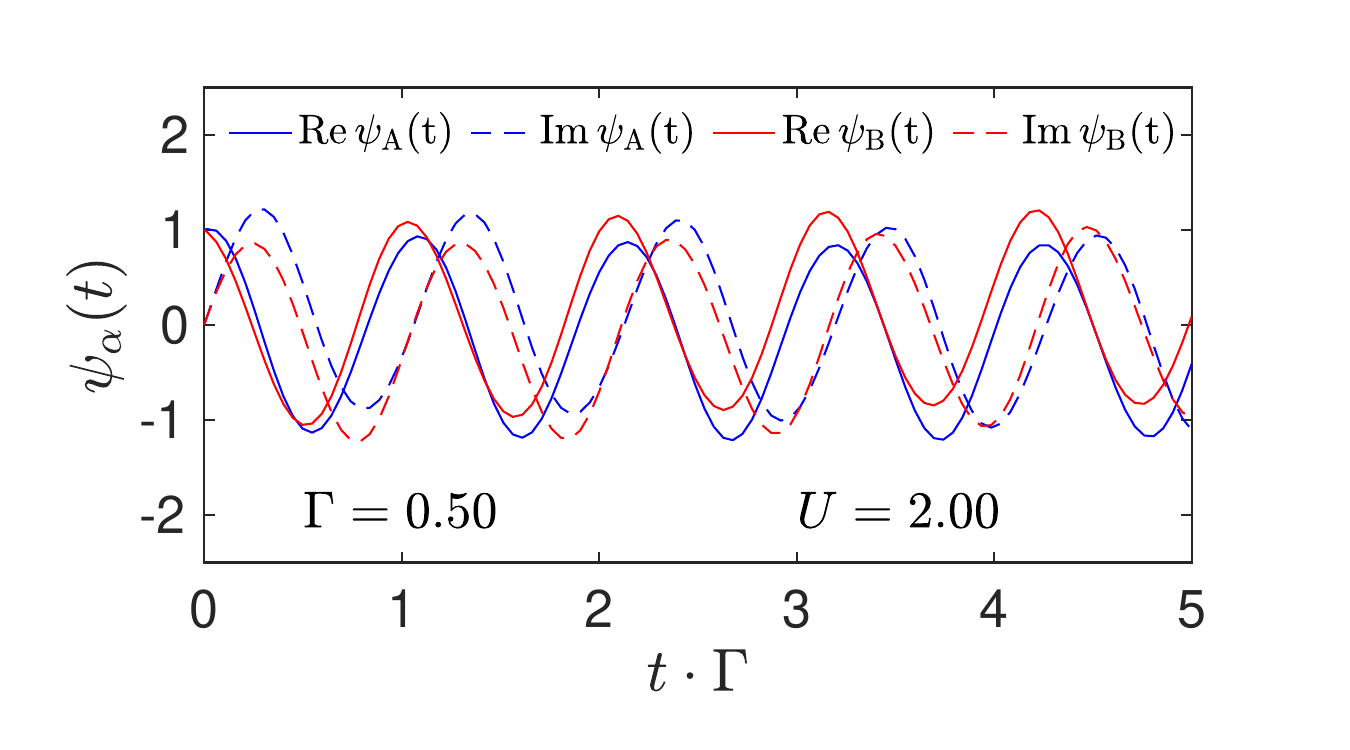}
	\includegraphics[width=0.33\columnwidth]{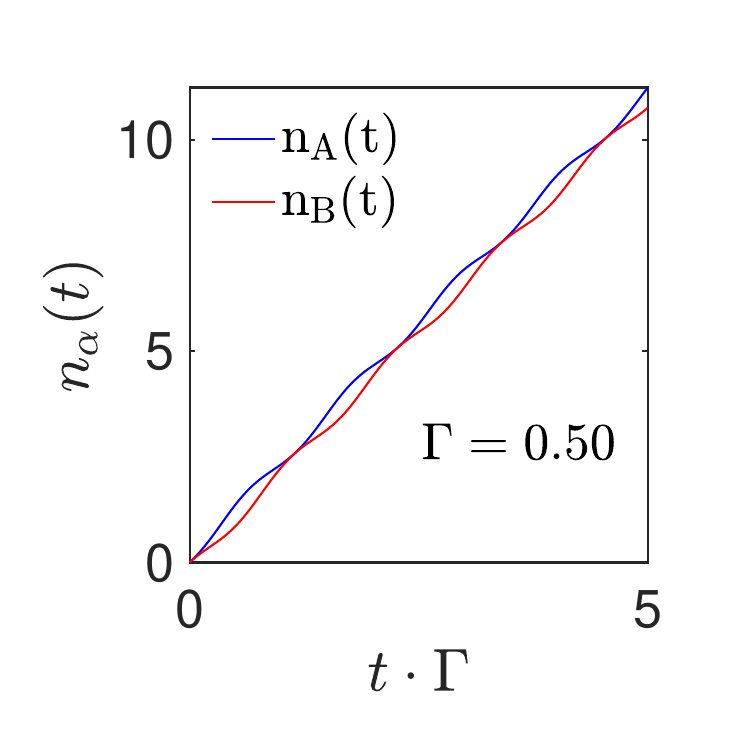}	
	\includegraphics[width=0.6\columnwidth]{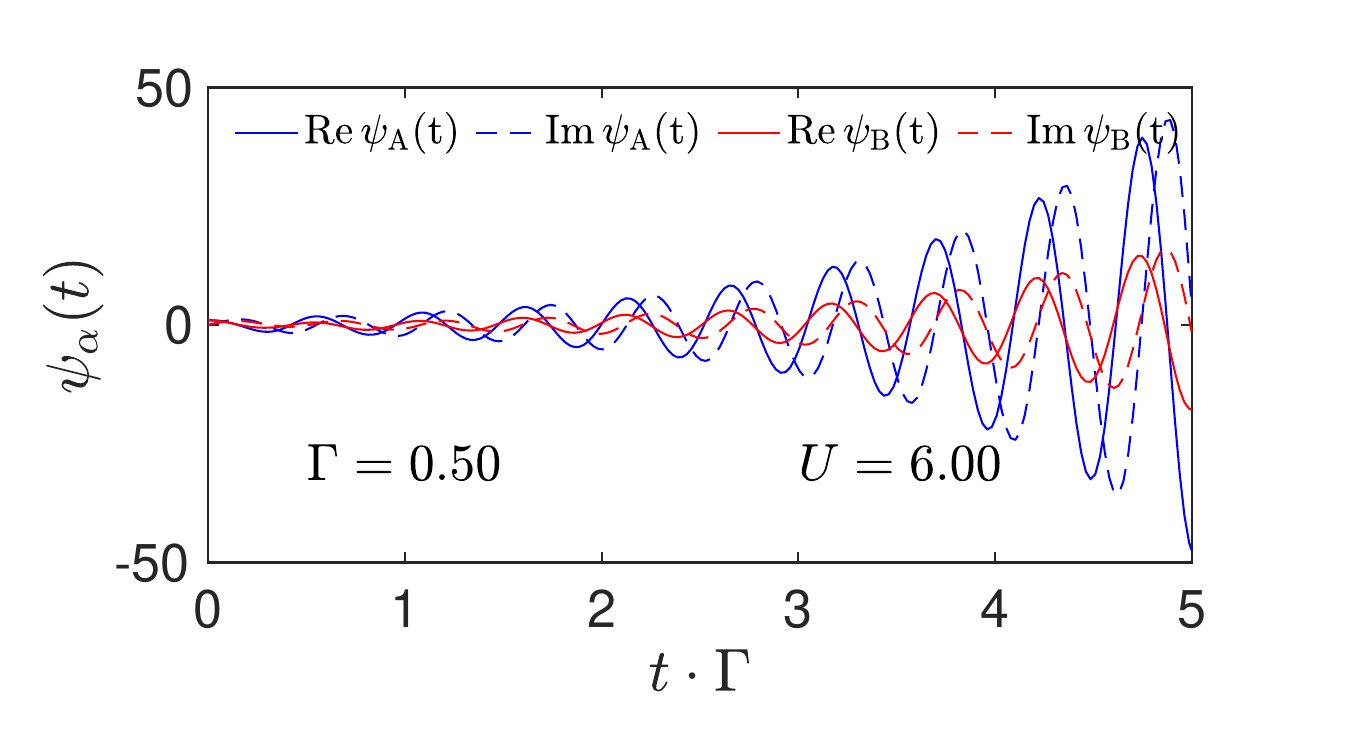}
	\includegraphics[width=0.33\columnwidth]{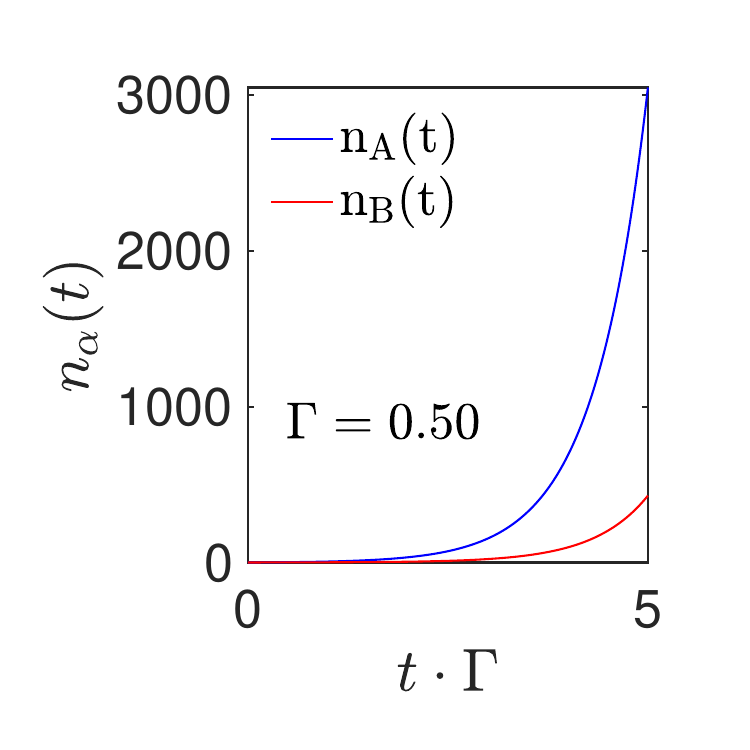}	
	\includegraphics[width=0.6\columnwidth]{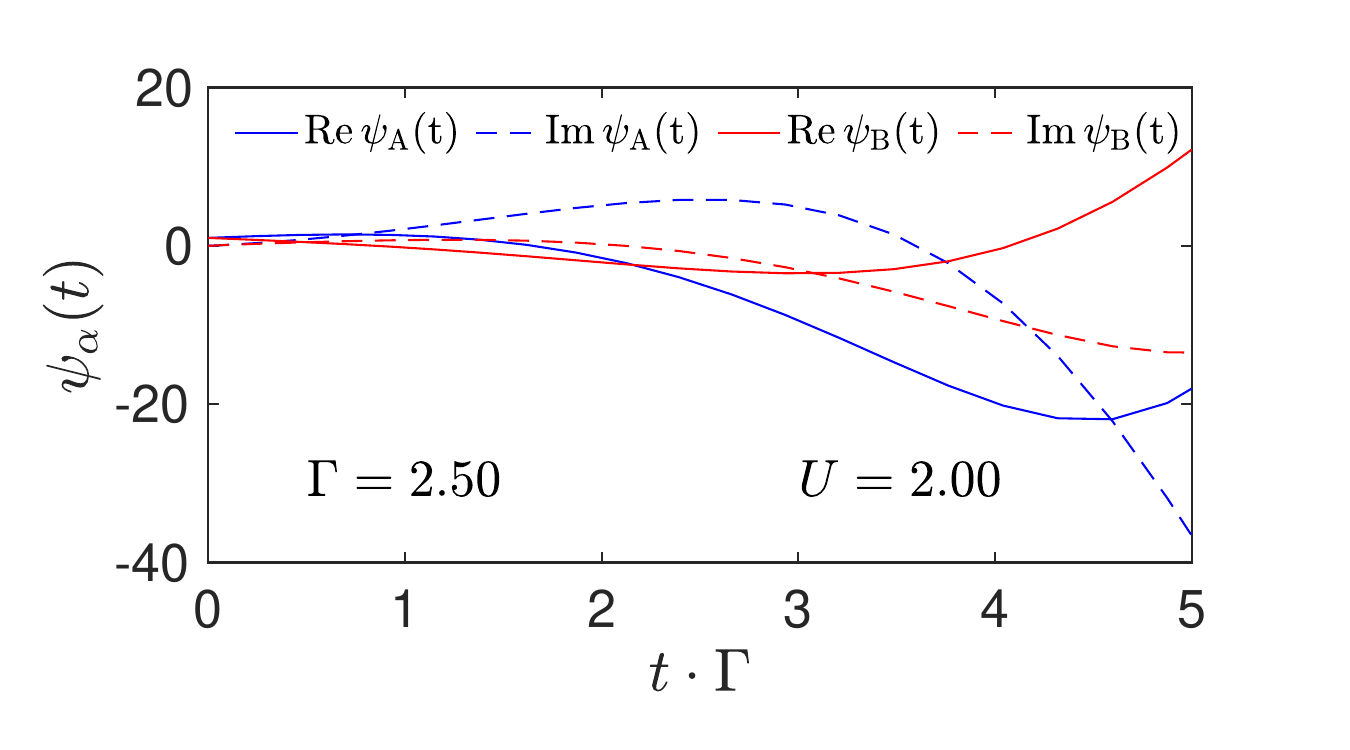}
	\includegraphics[width=0.33\columnwidth]{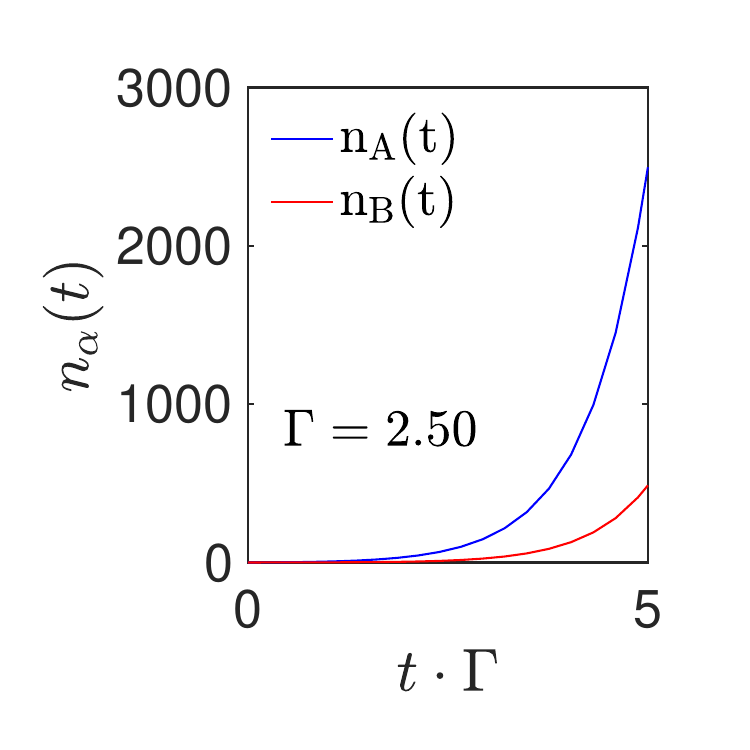}	
	\includegraphics[width=0.6\columnwidth]{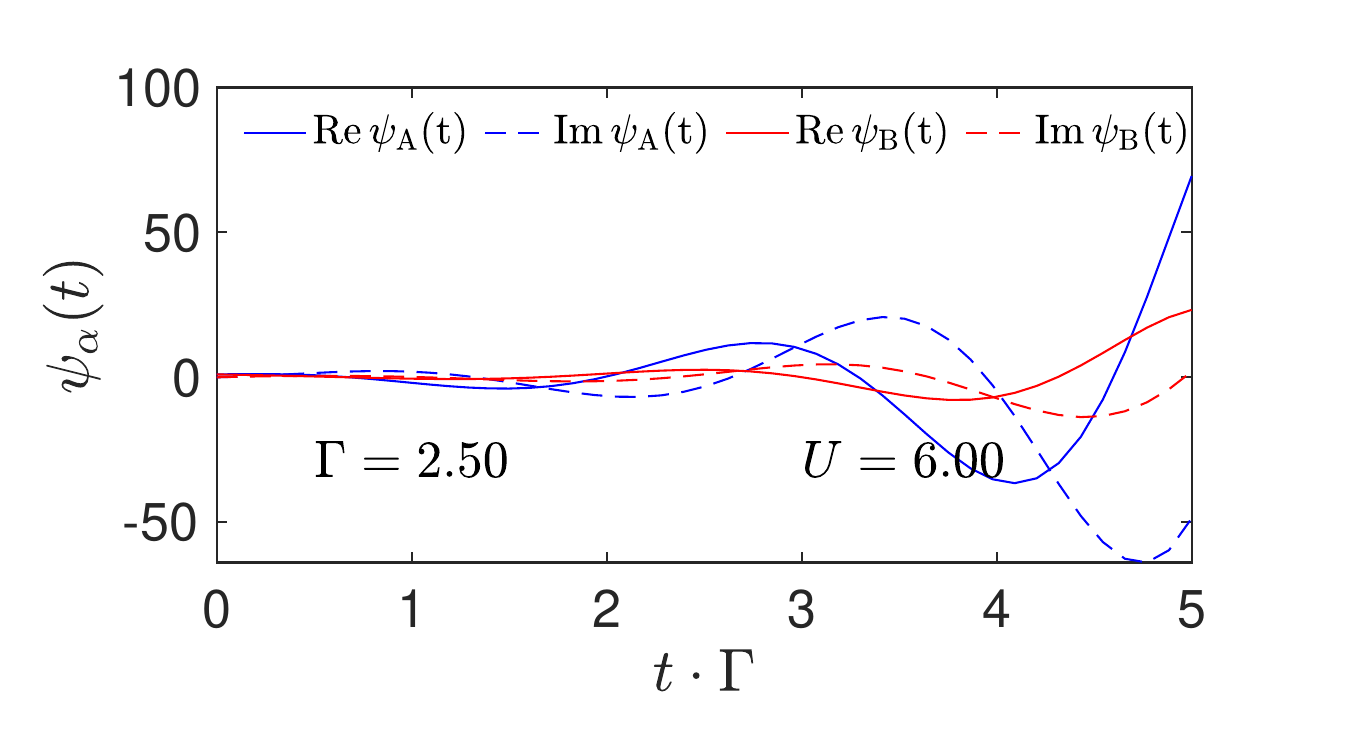}
	\includegraphics[width=0.33\columnwidth]{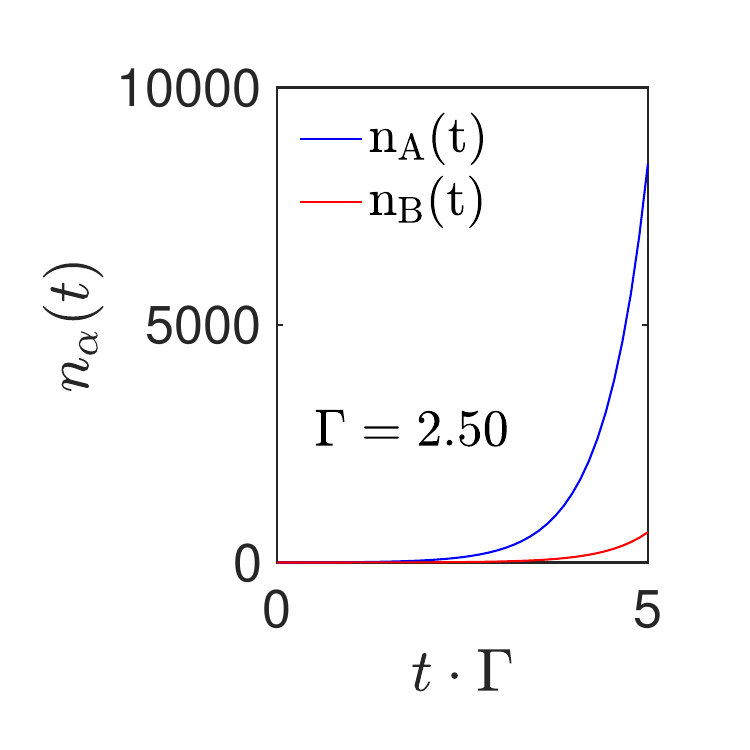}		
	\caption{The real and imaginary parts of the numerical solution for Eqs.~\eqref{eq:GPE} 
	for a finite $U$. Because of the cubic powers, the solutions are always complex. 
	In the long-time limit, the solutions remain periodic irrespective of $U$ or $\Gamma$ strength, but their amplitude grows in time.}
	\label{fig:comparison_psi_AB_U}
\end{figure}
In the ${\cPT}$-symmetric regime, corresponding to $\Gamma<\Gamma_C$, the wave functions are either real or imaginary, and are bounded and periodical. Their oscillation in time is associated to Rabi oscillations between neighboring sites. 
According to Eq.~\eqref{eq:n_i}, the cumulative integration gives the occupation at each site.  Apart from an oscillating envelope, the occupation displays a linear increase in time $\average{n_\alpha(t)}\propto t $.
The balanced gain and loss ensure an equally populated level, $\average{n_A(t)} \approx \average{n_B(t)}$ in the long-time limit. 
Correspondingly, the population imbalance that we have associated with the order parameter of the transition vanishes, $\Delta(\Gamma <\Gamma_C) =0$. 
Approaching the EP from the $\cPT$-symmetric side, the period of the oscillation diverges.
When $\Gamma>\Gamma_C$, the $\cPT$ symmetry is broken and the oscillatory behavior in the wave functions turns into an exponential dependence as function $t$, and diverges in the $t\to\infty$ limit. 
The balance between gain and loss is also broken (see Fig.~\ref{fig:comparison_S_AB}) 
and a finite order parameter $\Delta(\Gamma>\Gamma_C)$ develops with an amplitude that depends only on $\Gamma$. 
Using Eq.~\eqref{eq:n_i} and computing the integrals yields the order parameter $\Delta$ in the $t\to\infty$ limit:
\begin{equation}
\Delta = \left \{
\begin{array}{lr}
0 & \text{for}\, \Gamma<\Gamma_C,\\
\sqrt{1-\left (  \frac{\Gamma_C}{\Gamma} \right )^2 } & \text{for}\,  \Gamma>\Gamma_C.
\end{array}
\right .
\label{eq:Delta}
\end{equation}
This result, although quite simple, is nicely corroborated by the direct numerical evaluation of $\Delta$. 
In Fig.~\ref{fig:comparison_Delta}, we present the evolution of the order parameter in the two regimes. 
The symbols are obtained by direct numerical integration while the solid lines represent the analytical result derived in Eq.~\eqref{eq:Delta}. 
The phase change at the EP has all the characteristics of a second-order phase transition, 
while taking place in nonequilibrium conditions. Close to the transition point, the scaling of the order parameter displays a typical critical behavior,
$\Delta(\Gamma>\Gamma_C)\approx (\Gamma-\Gamma_C)^\alpha$ with an exponent $\alpha = {1\over 2} $ (see the inset in Fig.~\ref{fig:comparison_Delta}). 

The presence of the critical point is manifested also in the time evolution of the thermodynamic entropy.
For a quadratic Hamiltonian coupled to the environment, it is possible to determine the 
thermodynamic entropy exactly from the eigenvalues $\zeta_k$ of the 
correlation matrix $\sigma_{ij}^{\alpha\beta}(t)$~\eqref{corr_mat}
as in Ref.~\cite{Peschel2009},
\begin{equation}
S(t) = \sum_k (1+\zeta_{k})\ln(1+\zeta_k)-\zeta_k\ln\zeta_k.
\end{equation}
The entropy follows readily using Eqs.~\eqref{eq:S_AB_cos}, \eqref{eq:S_AB_cosh}, and \eqref{corr_mat}.
We find that $S$ scales differently in the two phases, and in the long-time limit, $t\Gamma\gg 1$,
changes from a logarithmic behavior $S(t)\sim \ln t$ in the $\cPT$-symmetric region to a linear 
growth $S(t)\sim t$ in the $\cPT$-broken phase, 
\begin{equation}
S(t)
\sim
\begin{cases}
2\ln({t\Gamma\Gamma_C \over \Gamma^<_{\rm eff}}) +2 \ln L  & \text{ for }\Gamma<\Gamma_C,\\
2t\Gamma^>_{\rm eff} +2\ln L
& \text{ for }\Gamma>\Gamma_C,
\end{cases}
\end{equation}
with $L$, the number of unit cells in the system.
These asymptotic results are corroborated in Fig.~\ref{fig:entropy} with the numerical 
results.
Exactly at $\Gamma=\Gamma_C$ we get $S(t)\sim 3\ln(\Gamma_Ct)+2\ln L $.

\section{The role of interactions, \texorpdfstring{$U\ne 0$}{U!=0}}\label{sec:U_ne_0}

\begin{figure}[t!]
	\includegraphics[width=0.95\columnwidth]{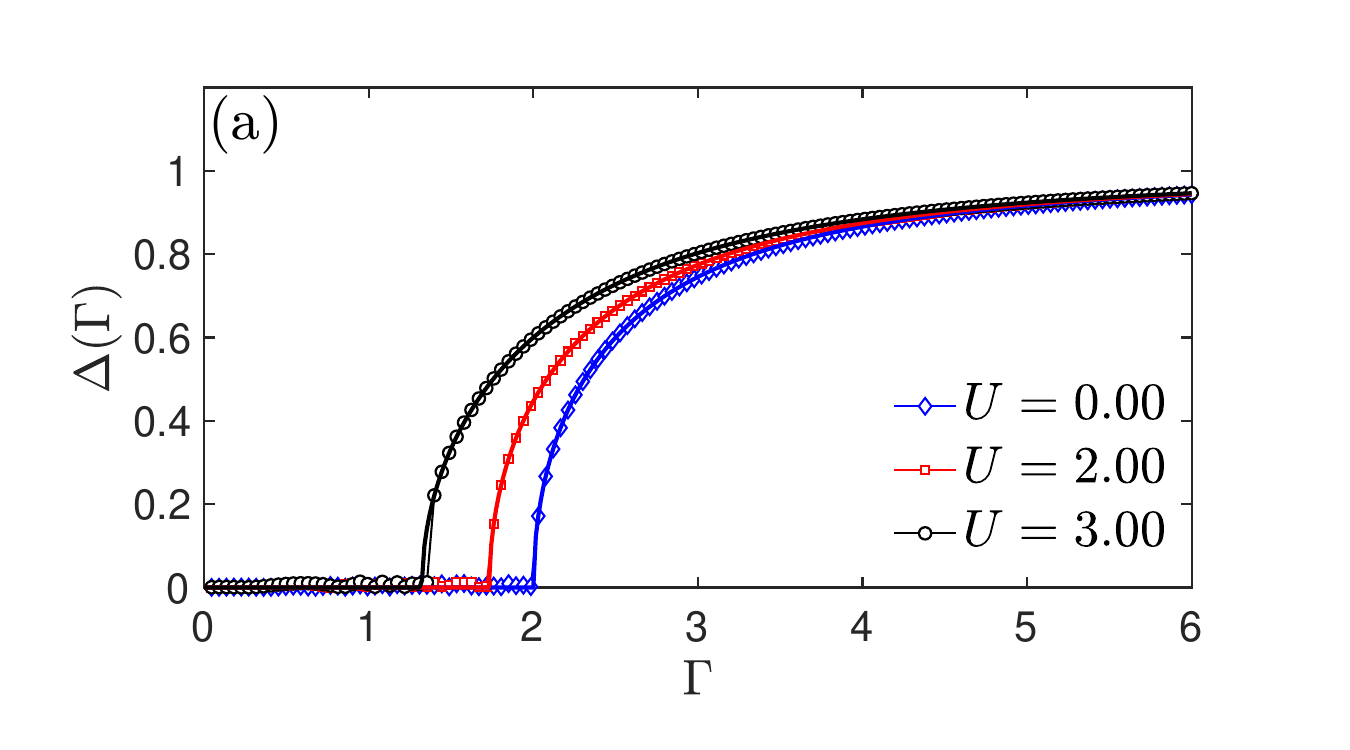}		
	\includegraphics[width=0.95\columnwidth]{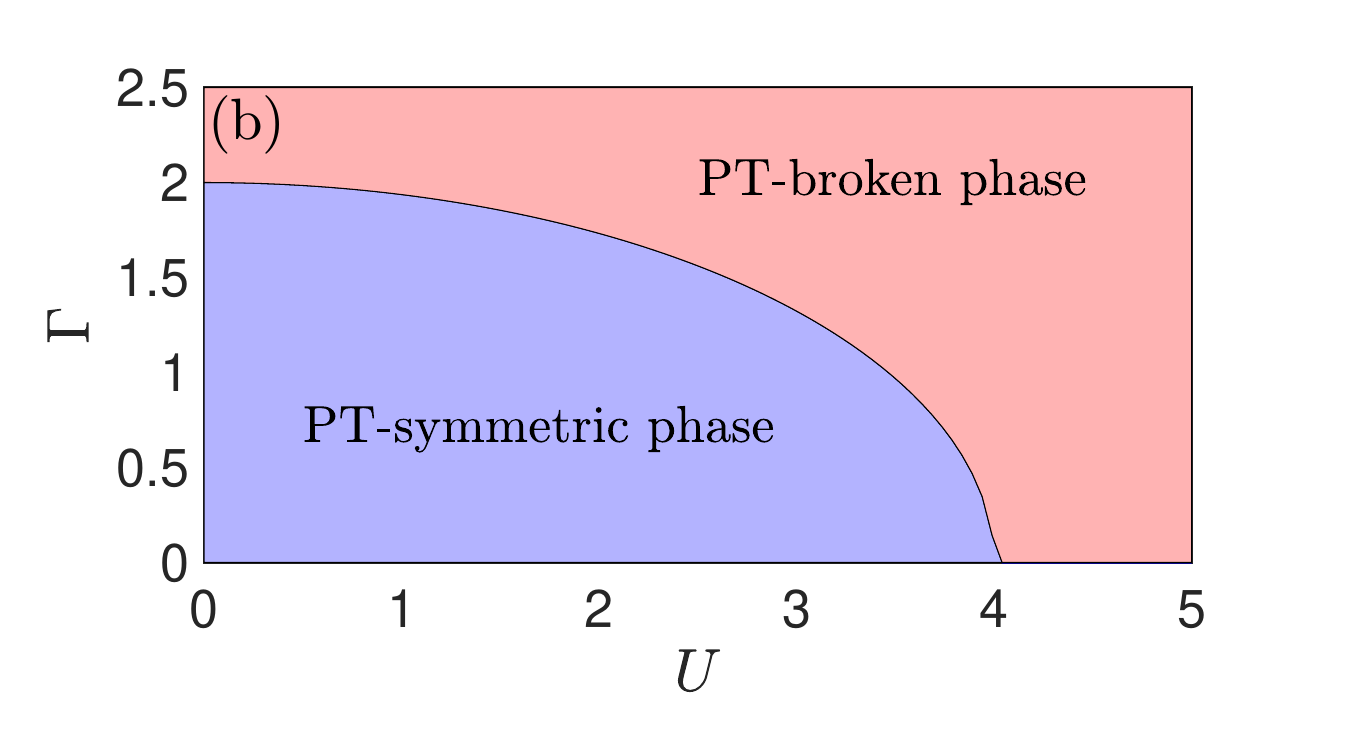}		
	\caption{ (a) Order parameter $\Delta(\Gamma)$  as function of  $\Gamma$ for different values of the interaction strength $U$. The solid lines are the analytical results. The plot 
	corresponding to $U=0$ is added for completeness.
	(b) Mean-field phase diagram  in the $(\Gamma, U)$ parameter space. 
	The separation line corresponds to the equation $\Gamma_C(U) = \sqrt{\Gamma_C^2-(U/2)^2}$.
	}
	\label{fig:phase_diagram}
\end{figure}

When the interaction is turned on, the set of equations \eqref{eq:psi_general} becomes nonlinear and there is no analytical solution. Still,  the connection to the non-Hermitian two-level system 
discussed in Sec.~\ref{sec:U=0} remains valid because the system is translation invariant.
When PBC are used, solving Eq.~\eqref{eq:psi_general} reduces to the GPE for a simple, 
double-well potential with symmetric gain and loss described by a nonlinear Hamiltonian.

We are interested in the limit when the number of particles in the system is infinite, 
at which point it can be characterized by a single macroscopic wave function~\cite{Dast2014}, which 
requires a proper normalization~\cite{Graefe2008}. Taking into account the norm $N(t)=|\psi_A|^2+|\psi_B|^2$,
which diverges as well in the long-time limit, i.e., by considering $\psi_{A,B}\over \sqrt{N(t)}$, the GPE gets modified as~\cite{Graefe2012}
%
%
\begin{equation}
H^{\rm eff}_{\rm int}(\psi_A, \psi_B)=\left (
\begin{array}{cc}
 U{|\psi_A|^2\over N(t)}+ i \Gamma & -2J\\
-2J &  U{|\psi_B|^2\over N(t)}-i\Gamma 
\end{array}
\right ).
\label{eq:H_int}
\end{equation}
In terms of the original fields describing the lattice model, the modified GPE reads 
\begin{eqnarray}\label{eq:GPE}
\dot \psi_{A, n} (t) &= &iJ \big( \psi_{B, n-1}+\psi_{B, n}  \big) +\Gamma \psi_{A, n}\nonumber \\ 
& &{} +i U {|\psi_{A, n}|^2\over N_n } \psi_{\psi, n},\nonumber\\
\dot \psi_{B, n} (t) &= &iJ \big( \psi_{A, n}+\psi_{A, n+1}  \big) -\Gamma \psi_{B, n}\\ 
& &{}+i U {|\psi_{B, n}|^2\over N_n} \psi_{B, n}\nonumber. 
\end{eqnarray}
Because of the cubic terms, the solution of~\eqref{eq:GPE} for the fields $\psi_{\alpha}(t)$ 
has  always  complex solutions. We solve the set of equations~\eqref{eq:GPE} numerically. 
Some typical result for different values of $\Gamma$ and $U$ are presented in Fig.~\ref{fig:comparison_psi_AB_U} 
together with the time evolution for the site occupation. 
\begin{figure}[t!]
	\includegraphics[width=0.6\columnwidth]{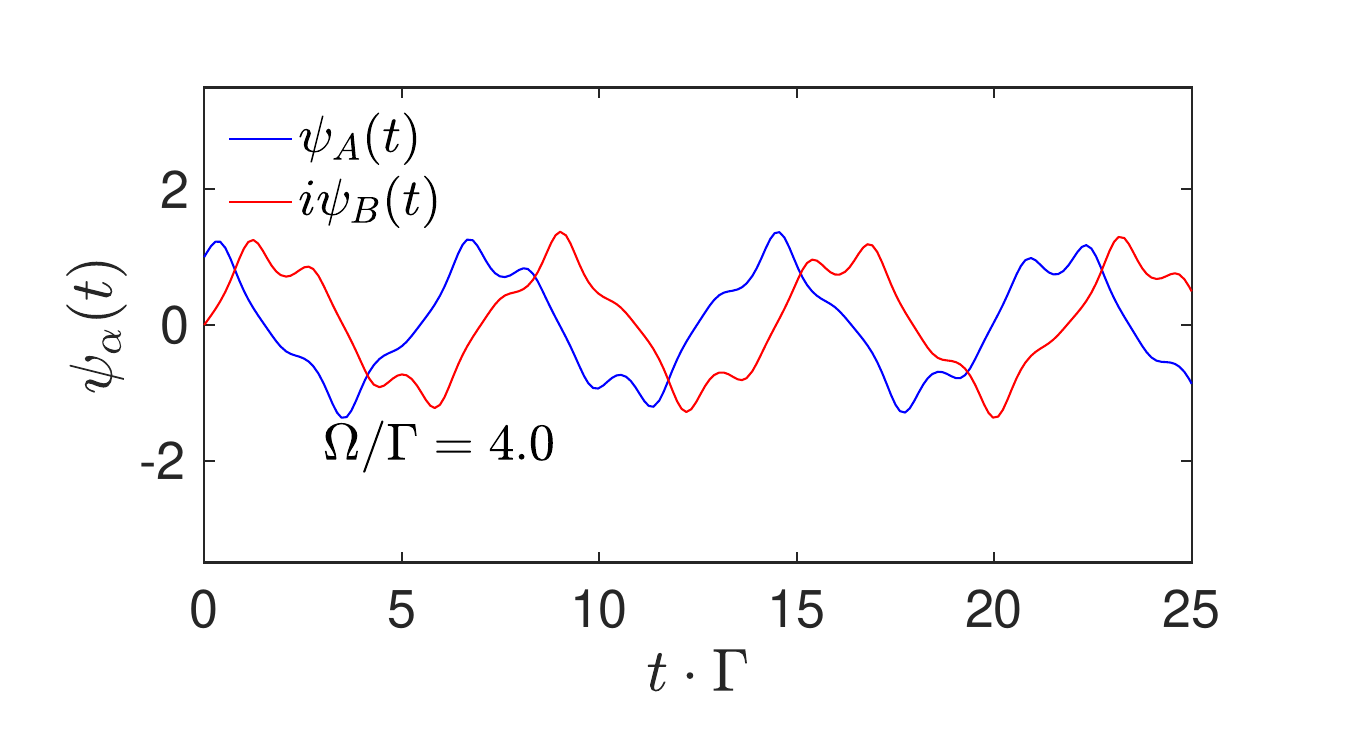}
	\includegraphics[width=0.33\columnwidth]{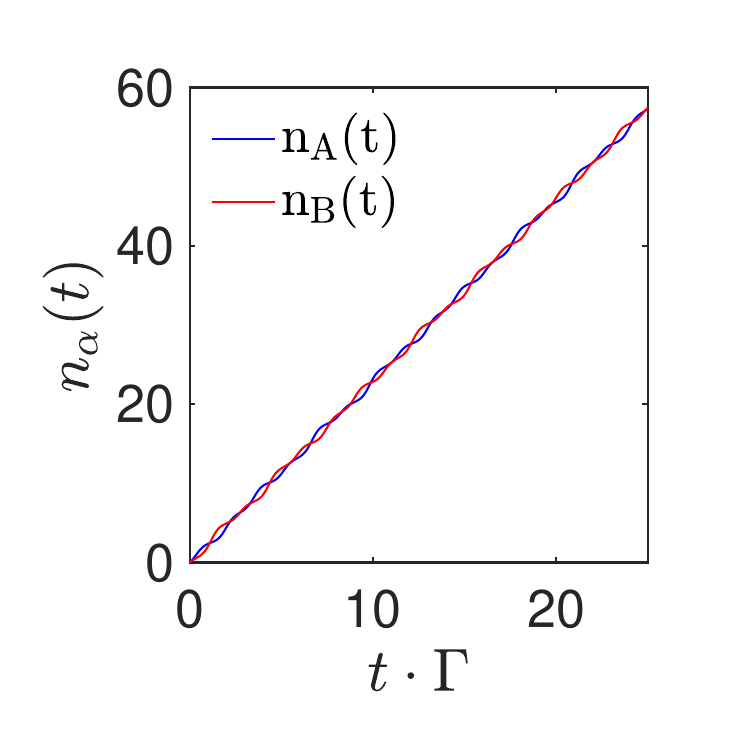}	
	\includegraphics[width=0.6\columnwidth]{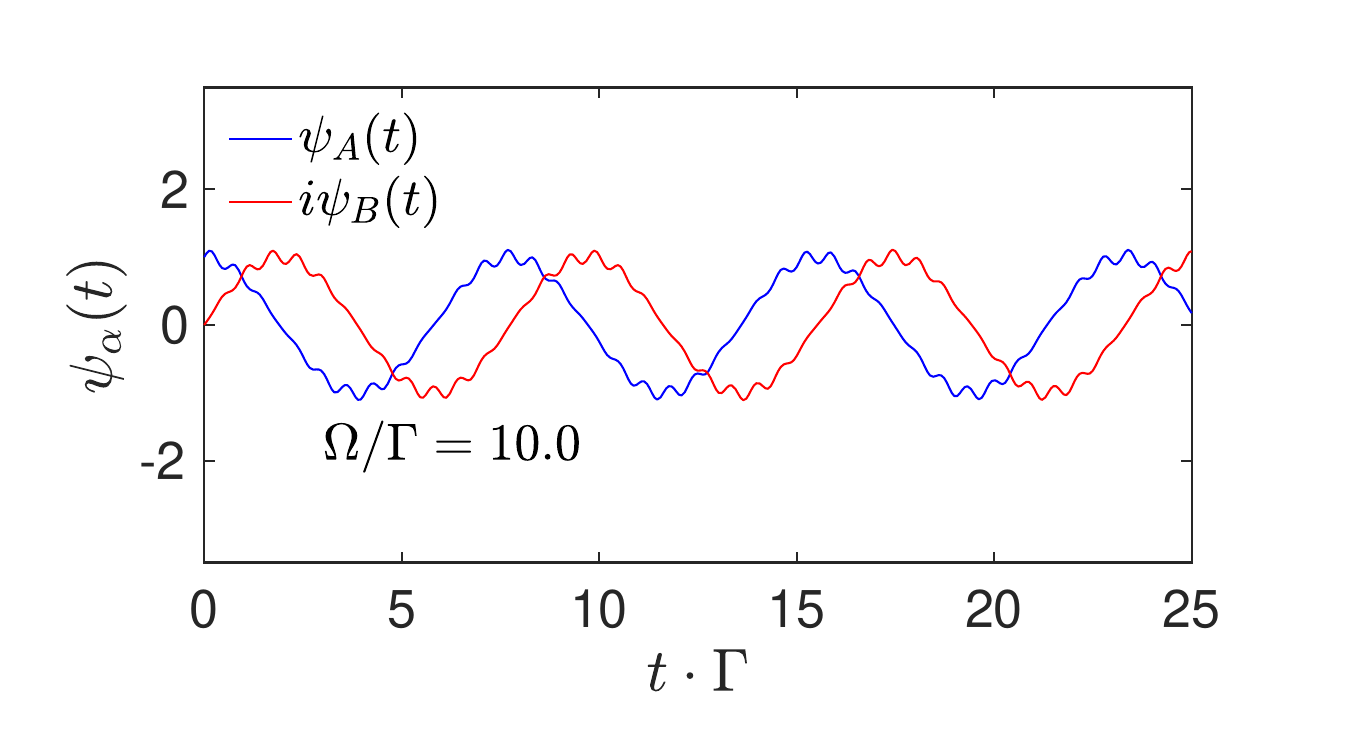}
	\includegraphics[width=0.33\columnwidth]{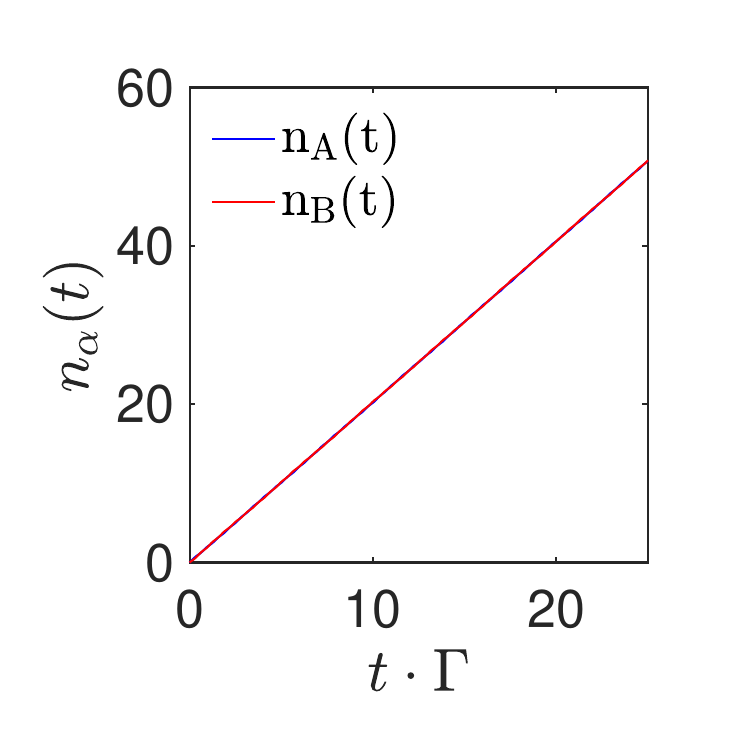}	
	\caption{The real and imaginary parts of the numerical solution for Eqs.~\eqref{eq:GPE} for a large driving frequency $\Omega$. When $\Omega\gg \Gamma$, irrespective of the $\Gamma$ value, the order parameter vanishes, and the system remains in the $\cPT$-symmetric regime. Correspondingly the occupation grows linearly in time.  The strength of the interaction is fixed to $U=0$, while $\Gamma=2.5$ in both cases. In the absence of interactions, the wave functions are either real or imaginary. }
	\label{fig:comparison_psi_AB_U_periodic}
\end{figure}

Further insight is gained by analyzing the stationary solutions in the 
long-time limit, using the ansatz $\psi_{\alpha} (t) = \sqrt{N_\alpha(t)} S_\alpha e^{-i\mu t}$, with normalized wave functions $|S_A|^2+|S_B|^2=1$, and $\mu$, a generally complex eigenvalue of the stationary GPE.
From the stationary problem, we compute the population imbalance and correspondingly the order parameter $\Delta=|S_A|^2-|S_B|^2$.
There are always solutions for $\Delta=0$ that persist from the linear noninteracting problem, as one can check directly in Eq.~\eqref{eq:GPE}, since the nonlinearity drops out for $\Delta=0$, and $\mu=\pm\sqrt{\Gamma_C^2-\Gamma^2}$.
However, these solutions do not break the $\cPT$ symmetry for $\Gamma>\Gamma_C$, but simply vanish at the critical $\Gamma_C$~\cite{Graefe2006,Graefe2012}.
Instead, the $\cPT$-symmetry breaking is due to specific solutions of the interacting problem, for finite real $\Delta$: $\mu(U)=U/2+i\Gamma \Delta$.
Solving Eq.~\eqref{eq:GPE} with $\mu(U)$ for $\Delta\geq 0$ yields the solution
\begin{equation}
\Delta =
\begin{cases}
0 & \text{for}\, \Gamma<\sqrt{\Gamma_C^2 -\big ({U\over 2}\big )^2},\\
\sqrt{1-\frac{\Gamma_C^2}{\Gamma^2+(U/2)^2 }} & \text{for}\,  \Gamma>\sqrt{\Gamma_C^2 -\big ({U\over 2}\big)^2},
\end{cases}
\label{eq:Delta_U}
\end{equation}
where $\Delta=0$ indicates that only trivial solutions from the $U=0$ case survive.
Thus, the position of the critical point where $\cPT$-symmetry gets broken, and therefore an imbalance between site $A$ and $B$ populations starts to develop, is shifted by interactions, $\Gamma_C\to\Gamma_C(U)=\sqrt{\Gamma_C^2 -(U/2)^2}$. The results for $\Delta$ in Eq.~\eqref{eq:Delta_U} match
perfectly the numerical ones obtained by integrating 
directly Eq.~\eqref{eq:GPE} [see Fig.~\eqref{fig:phase_diagram}(a)].
Eq.~\eqref{eq:Delta_U} indicates that the critical exponent $\alpha = 0.5$ remains the same in the presence of interactions.  
In Fig.~\ref{fig:phase_diagram}(b), we represent the 
phase diagram of the model. In a finite region in the parameter space ($\Gamma, U$), the system remains in the 
$\cPT$-symmetric regime. For large enough interaction strengths $U$ or for large dissipation $\Gamma$.
the system is always in the $\cPT$-broken phase.   

Finally,  we briefly discuss the problem of periodic driving~\cite{Kohler1997, Sieberer2013, Chitsazi2017, Schnell2020,  Ikeda2021}
in which the dissipation is modulated periodically in time, what we formally write as $\Gamma(t) = \Gamma\cos(\Omega\, t)$.
In the original Lindbladian,  a negative $\Gamma$ would correspond to a site with losses, while a positive one to a site where gains occur, 
so modulating $\Gamma(t)$  corresponds to a periodic switching of the sites with gain and loss. 
In Eqs. ~\eqref{eq:GPE}, it simply translates to replacing the constant $\Gamma$ with $\Gamma(t)$.
It is obvious that the driving will have a strong effect on the population dynamics. 
The static limit, $\Omega\to 0$ would correspond to the analysis that we performed so far. 
Another limit of interest is that of strong driving, $\Omega\gg \Gamma$, in which case the fast 
exchange between the sites with gain and loss, is easily understood as well. Because of the fast alternation 
of the gain-loss sites, the wave function $\psi_\alpha(t)$ has an oscillating behavior as well, is bounded, and 
implicitly the imbalance between the particles at the gain-loss sites vanishes in the long-time limit. 
This is displayed in Fig.~\ref{fig:comparison_psi_AB_U_periodic}.
Basically, in the strong-driving limit, the critical coupling is pushed to higher values of the order $\Gamma_C\approx \Omega$.  
We have numerically checked that for any finite $\Omega$, adding a finite interaction $U$ does not change the physics, but only shifts the transition line in Fig.~\ref{fig:phase_diagram} to larger values of both $\Gamma$ and $U$.  

\section{Conclusions}\label{sec:Conclusions}
We examined the dissipative dynamics of a one-dimensional bosonic system that exhibits $\cPT$ symmetry under certain conditions and displays a $\cPT$-symmetry phase transition that is characterized by an order parameter, much like continuous phase transitions in Hermitian systems. 
The particular model that we consider is the bipartite Bose-Hubbard model with alternating gain and loss. 
We determine the order parameter in the noninteracting limit while in the presence of interaction 
the problem is addressed at the mean-field level. This allows us to construct the phase diagram of the model.
We find that increasing both the interaction and dissipation rates induce a $\cPT$-symmetry breaking.
In addition, by periodically modulating the dissipative coupling in time stabilizes the $\cPT$-symmetric regime and washes out the chances of $\cPT$-symmetry breaking.
In the $\cPT$-symmetric regime, the thermodynamic entropy increases as $S(t)\sim \ln t$, while in the $\cPT$-broken phase, it grows linearly in time, $S(t) \sim t$.

\section*{Acknowledgments}

This research is supported by the National Research, 
Development and Innovation Office  NKFIH within the Quantum Technology 
National Excellence Program (Project No. 2017-1.2.1-NKP-2017-00001), No.~K142179, No.~K134437, No.~SNN139581,
by the BME-Nanotechnology FIKP grant (BME FIKP-NAT), and by a grant of the Ministry of 
Research, Innovation and Digitization, CNCS/CCCDI-UEFISCDI, under Project No.~PN-III-P4-ID-PCE-2020-0277, under the project for funding the excellence, 
Contract No.~29 PFE/30.12.2021, and ``Nucleu'' Program 27N/03.01.2023, Project No.~PN 23 24 01 04.
\bibliography{references}

\end{document}